\documentclass[aps,twocolumn,prl]{revtex4}
\usepackage{graphicx,subfigure}
\begin{document}
\title{Josephson junctions as threshold detectors for full counting statistics}
\author{J. Tobiska and Yu. V. Nazarov}
\affiliation{Kavli Institute of NanoScience, Delft University of Technology,
2628 CJ Delft, the Netherlands}
\date{\today}
\begin{abstract}
We discuss how threshold detectors can be  used for a direct measurement of the full distribution of current fluctuations and how to exploit Josephson junctions in this respect. We propose a scheme to characterize the full counting statistics (FCS) from the current dependence of the escape rate measured. We illustrate the scheme with explicit results for tunnel, diffusive and quasi-ballistic mesoscopic conductors. 
\end{abstract}
\maketitle

Quantum noise in electron transport is an actively developing field.
Noise measurements provide exclusive information about microscopic mechanisms of the transport that can hardly be obtained by other means~\cite{shotnoise,qnoise}. Still, the experiments in the field  neither match
the intensive theoretical development nor gather {\it all} information about electric fluctuation. Indeed, the concept of full counting statistics pioneered in~\cite{jmphyslevles} allows one to predict the non-Gaussian distribution
function of the current measured during a time interval $\tau$, $P_{\tau}(I)$. This distribution is characterized by an infinite set of cumulants $\ll I^{n}\gg $. A traditional noise measurement only assesses the second cumulant
of this set discarding the rest. Recent pioneering work reports a successful measurement of the third cumulant~\cite{thrdcumulant}, but there is a long way to go if one measured the cumulants one by one. It would be advantageous to measure the distribution function directly and thus to  get all cumulants at once, thereby collecting the wealth of information being currently
discarded.

Why is such a measurement difficult? The probabilities to measure correspond to big deviations of the current from its average value, $|I -\langle I \rangle | \simeq \langle I \rangle$, and are therefore exponentially small. For instance, in the shot noise regime $P_{\tau}(I) \simeq \exp \left( - \langle I \rangle {\cal G}(I/\langle I \rangle)\tau/e\right)$, ${\cal G}(I/\langle I
\rangle) \simeq 1$ being the function to characterize. One has to concentrate on very rare measurement outcomes that occur with probability $\exp(-\langle I \rangle\tau/e) \approx 0$. Such measurements can only be carried out with {\it threshold detectors} that discriminate these rare events. 
Let us discuss an {\it ideal} threshold detector that measures the current during the time interval $\tau$, and gives a signal if the current
measured exceeds the threshold current $I_\mathrm{th}$. The signal probability will then be proportional to $P_{\tau}(I_\mathrm{th})$. To give a realistic illustration, a detector that measures a tunnel junction with $\langle I \rangle = 10~\mbox{pA}$ in the time interval $\tau = 10^{-6} \mbox{s}$ would go off once
an hour if $I_\mathrm{th} = 2 \langle I \rangle$ and once in $10^{-4} \mbox{s}$ if $I_\mathrm{th} = 1.5 \langle I \rangle$. Therefore, if one measures the rate of the detector signals as a function of $I_\mathrm{th}$, one directly assesses the full counting statistics.

Albeit realistic detectors are not ideal. There are three important factors that can either hinder the interpretation of such a measurement or even prevent the measurement: (i) a realistic detector hardly measures the current averaged over a certain time interval $\tau$. It is {\it dispersive}, being usually more sensitive to longer and smaller current fluctuations rather than to bigger and
shorter ones. (ii)  The detector may produce a significant
feedback on the system measured when it starts to signal, thereby disrupting its noise properties. (iii) The detector could just go off
by itself, for instance, due to quantum tunneling.

A Josephson junction seems to be a natural threshold detector for current fluctuations. It can be viewed as a particle in a washboard potential~\cite{tinkham}, the superconducting phase difference $\phi$ across the junction corresponding to the particle's coordinate. The junction is in zero-voltage state provided the current does not exceed the critical value corresponding to the critical tilt of the washboard potential. $\phi$ is trapped in one of the minima of the potential, which is separated by a barrier from the neighboring one. A current fluctuation that exceeds the critical threshold sets $\phi$ into motion and the junction gives a signal---a voltage pulse that lasts till $\phi$ is retrapped in a different minimum.

In this paper we address the feasibility of Josephson junction systems for measuring the full distribution of current fluctuations in a mesoscopic conductor. Our results are as follows. The Josephson junction is a realistic detector, all three factors mentioned are in play. Albeit one can measure the distribution provided the width of the barrier $\phi_0 \gg 1$. This can be realized by a flux division using two inductances.
 Under these conditions, the third factor is of no importance and the first and second factor do not hinder the unambiguous correspondence between FCS and the escape rate of the junction as a function of $I_\mathrm{th}$. These theoretical results open the way to direct experimental observation of FCS.\@

The circuit under consideration consists of a normal coherent conductor with conductance $G$ in series with the Josephson junction(system) (Fig. 1). The system is biased with voltage source $V \gg k_\mathrm{B} T/e$. This assures that the normal conductor is in the shot noise regime and its actual temperature is not relevant. In addition, we inject extra current $I_\mathrm{b}$ that controls the slope of the Josephson washboard potential. 

\begin{figure}
  \includegraphics[width=0.9\columnwidth]{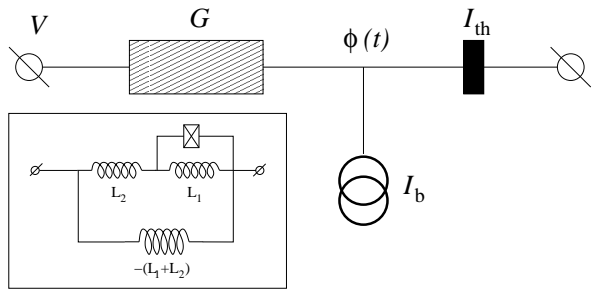}
  \caption{A voltage biased mesoscopic conductor with conductance $G$ provides the noise source for a threshold detector which is characterized by its threshold current $I_\mathrm{th}$. $I_\mathrm{b}$ is an additional current bias. The inset shows a possible realization of the detector with two inductances $L_{1,2}$, a Josephson junction and a negative inductance element.}
\label{fig:system}
\end{figure}

If fluctuations are neglected, this system can be described with the celebrated model of resistively shunted junction~\cite{tinkham}. The normal conductor is a source of non-Gaussian current fluctuations that instantly tilt the washboard potential and can lead to an escape of $\phi$ from the minimum. The escape gives rise to an observable voltage pulse. The escape rate in the same or similar systems has been studied for a variety of noise sources and potentials~\cite{kramers40,fultondunkleberger74,buettharrland83,larkin}. To our knowlegde, the non-Gaussian noise sources that are characterized by FCS were not adressed yet. 

To proceed, we begin with the fully quantum mechanical description of the system in terms of a Keldysh action for a single variable $\phi$~\cite{schoen,larkin}, that incorporates information about FCS of the normal conductor and the properties of the Josephson junction. We calculate the escape rate by considering saddle-point trajectories of the action, $A$, that connect the potential minimum with the nearest potential maximum. With exponential accuracy, the rate is given by $\Gamma \simeq \exp(-\mathrm{Im} A/\hbar)$.

The action consists of two terms, $A=A_\mathrm{J}+A_\mathrm{N}$, corresponding to the elements of the circuit. We denote by $\phi^\pm$ the phases on the forward/backward parts of the Keldysh contour and also use symmetrized combinations of these
$\phi,\chi=(\phi^+\pm\phi^-)/2$. The junction part reads in a standard way~\cite{schoen}:
\begin{equation}
  A_\mathrm{J}=\int dt\left(U(\phi^+(t)) - \frac{\hbar^2 C}{8 e^2}\dot{\phi}^{+2}(t)\right) - \{ \phi^+ \leftrightarrow \phi^-\},
\end{equation}
$C$ being the self-capacitance of the junction, $U(\phi)$ being the Josephson energy with the current bias term included: $-U(\phi)=(\hbar/2e)(I_\mathrm{c} \cos\phi +I_\mathrm{b} \phi)$ for a single junction. Further we concentrate on overdamped junctions where $C \ll G^2\hbar/(2e I_\mathrm{c}) $ and neglect the capacitance term. The normal conductor part we write following~\cite{kindermannnazarov} in quasi-stationary approximation 
\begin{equation} 
  A_\mathrm{N}=\frac{i\hbar}{2e}G\int dt (V - \frac{\hbar}{2e}\dot{\phi}(t))S(\chi(t));
\label{eq:actngeneral}
\end{equation}
where $S$ characterizes the FCS and the preceding factor is just the voltage drop over the normal conductor. The distribution of current noise is given by the Fourier transform of $S$. Derivatives with respect to $\chi$ generate the moments of the distribution.

A coherent conductor can be presented by a set of transmission eigenvalues $T_n$ and $S$ is given by Levitov's formula~\cite{jmphyslevles} 
\begin{equation}
S(\chi)=\frac{G_\mathrm{Q}}{G}\sum_n\ln
\left(1+T_n(e^{i\chi}-1)\right),
\label{eq:sgeneral}
\end{equation}
$G_\mathrm{Q}$ being the conductance quantum. Concrete forms of $S(\chi)$ for specific conductors will be given below. At $\chi \rightarrow 0$, $S$ can be expanded in $\chi$, $S \approx i\chi -\chi^2F/2$, $F$ being the Fano factor that describes the suppression of shot noise in comparison with the Poisson value~\cite{shotnoise}.

This quasi-stationary approximation is only valid if the typical time $\tau$ of the motion along the saddle-point trajectory is long in comparison with $\hbar /eV$, that is, $ eV\tau \gg \hbar$. To check the validity of this, we precede the results with simple  qualitative estimations. 

Let us consider an arbitrary barrier with the width $\phi_0$ and height $U_0 \simeq (\hbar/e) I_\mathrm{th} \phi_0$. The detection time can be estimated equating the potential energy term and the term with $\dot\phi$, $G \phi_0 {(\hbar/e)}^2 \chi /\tau \simeq U_0\chi/\phi_0$, $\chi$ being a typical value along the trajectory. This gives $\tau \simeq (\hbar/eV)\phi_0 (I_\mathrm{f}/I_\mathrm{th})$. The quasi-stationary approximation thus holds provided $I_\mathrm{f} \equiv GV \gg I_\mathrm{th}/\phi_0$. Let us estimate $\chi$ by equating the term which is quadratic in $\chi$ and the potential term. This gives $\chi \simeq I_\mathrm{th}/I_\mathrm{f}$ if $I_\mathrm{th} \ll I_\mathrm{f}$, $\chi \simeq 1$ otherwise. We see that if $\phi_0 \lesssim 1$ then $\chi \ll 1$. The latter implies that $S(\chi)$ can be expanded near $\chi=0$ and only the first two cumulants are relevant: no chance to see the effect of FCS.\@ However, if $\phi_0 \gg 1$, $\chi$ can become of the order of unity without violating the quasi-stationary approximation, and one can observe the FCS.\@ The quasi-stationary approximation remains valid for $\chi \lesssim \phi_0$.

The resulting rate can be estimated as  $\log \Gamma \simeq \phi_0 (G/G_\mathrm{Q})\chi$. If $\phi_0 \lesssim 1$, this reduces to  $\log \Gamma \simeq \phi_0 (G/G_\mathrm{Q})I_\mathrm{th}/I_\mathrm{f}$. In the opposite limit, the estimation for the rate reads $\log \Gamma \simeq \phi_0 (G/G_\mathrm{Q})\Xi(I_\mathrm{th}/I_\mathrm{f})$, $\Xi$ being a dimensionless function $\simeq 1$. It is important to note that these expressions match the quantum tunneling rate $\log \Gamma_{\hbar} \simeq U_0 \tau/\hbar \simeq (G/G_\mathrm{Q})\phi^2_0$ provided $eV\tau \simeq \hbar$. Therefore the quasi-stationary approximation is valid when the quantum tunneling rate is negligible and the third factor mentioned in the introduction is not relevant. For equilibrium systems, the situation corresponds to the well-known crossover between thermally activated and quantum processes at $k_\mathrm{B} T \tau \simeq \hbar$~\cite{larkin}.
 
We proceed with the quantitative solution. The trajectories we are looking at start at $t \rightarrow -\infty$ in the minimum of the potential with $\phi=\phi_\mathrm{min},\chi=0$ and approach the maximum  $\phi=\phi_\mathrm{max},\chi=0$ at $t \rightarrow \infty$. They obey the equations of motion
\begin{eqnarray}
  0&=&\frac{\partial}{\partial\chi}\left[U(\phi^+(t))-U(\phi^-(t))+\frac{i\hbar}{2e}G(V-\frac{\hbar}{2e}\dot{\phi}(t))S(\chi(t))\right],\\
  0&=&\frac{\partial}{\partial\phi}\left[U(\phi^+(t))-U(\phi^-(t))\right]+i{\left(\frac{\hbar}{2e}\right)}^2G\dot{\chi}\frac{\partial S}{\partial\chi}.
\end{eqnarray}
It is important to note that these equations have a simple integral of motion
\begin{equation}
  i(U(\phi^+)-U(\phi^-))+ \frac{\hbar}{2e}I_\mathrm{f} {S}(\chi)={\cal I}\label{eq:intmotion}
\end{equation}
${\cal I}=0$ for saddle-point trajectories of interest. The full action along the trajectory then reads
\begin{equation}
-\frac{2e^2}{\hbar^2G}A=\int dt \dot{\phi}S(\chi) =
\int_{\phi_\mathrm{min}}^{\phi_\mathrm{max}} d\phi S(\chi(\phi))
\label{eq:actnsp}
\end{equation}
where in the last relation $\chi$ is expressed in terms of $\phi$ by means of Eq.\ref{eq:intmotion}.

Let us start with the results for $\phi_0 \simeq 1$. In this case, one expands the action in terms of $\chi$ keeping terms of the first and second order only. This immediately yields $\chi = i4e (\partial U/\partial \phi)/(\hbar F I_\mathrm{f})$. The general answer for the escape rate can be obtained at any shape of the barrier and reads:
\begin{equation}
  \Gamma\simeq \exp \left( - \frac{U_\mathrm{max}-U_\mathrm{min}}{k_\mathrm{B} T^*}\right); \; \; 
  k_\mathrm{B} T^* = eV F/2
\label{eq:thermal}
\end{equation}
This is thermal activation with an effective temperature given by the noise in the normal conductor. A similar effect of noise was envisaged in a recent article~\cite{grabert} for the phase diffusion regime.


How to realize a device where the barrier width $\phi_0 \gg 1$?
It can not be just a single
Josephson junction since the phase drop on the junction
can not exceed $\pi$. We can make the phase drop 
over the junction much smaller than the phase 
drop over the device by flux division with two inductances
$L_{1,2}$ in series provided $\hbar/(e I_c) < L_1 \ll L_2$ (see inset Fig. 1).
However, this is not enough since the energy
of the device would be dominated by that of the inductances,
$\propto \phi^2/(L_2+L_1)$. This parabolic background
shall be compensated with a negative inductance $-(L_{1} +L_{2})$
in parallel. Such negative inductance can be made with the aid
of an active circuit \cite{sunegind,funatonegind} or properly biased
Josephson junction system \cite{semenovnegind}. This provides
a wide barrier $U(\phi)$.   

We notice that any function $U(\phi)$ can be approximated by a cubic parabola if the tilting of the washboard potential is close to the critical value. This is why we choose the  cubic parabola form 
\begin{equation}
  \frac{\partial U}{\partial \phi}=\frac{\hbar}{2e}I_\mathrm{th}\left[1-{\left(\frac{\phi}{\phi_0}\right)}^2\right],\label{eq:potappx}
\end{equation}
for actual calculations. It is convenient to require that the barrier does not change if we change $I_\mathrm{f}$. This can be done by a corresponding change of $I_\mathrm{b}$. To simplify this further, we notice that $\chi \ll \phi_0$ so that 
\begin{equation}
U(\phi^+)-U(\phi^-)\approx \chi\frac{\partial U}{\partial \phi}.\label{eq:potdrv}
\end{equation}
Substitution into Eq.~\ref{eq:intmotion} gives $\phi$ in terms of $\chi$  
\begin{equation}
  \phi=\phi_0\sqrt{1+\frac{I_\mathrm{f}}{I_\mathrm{th}}\left(\frac{S(\chi)}{i\chi}-1\right)}.\label{eq:phiofmu}
\label{eq:phi}
\end{equation}
Combining this with Eq.\ref{eq:actnsp}, we obtain the escape rates as a function of $I_\mathrm{th}/I_\mathrm{f}$ for any given FCS.\@

To stress similarities and differences with thermal activation, we present the results in the form of Arrhenius-like plots. We plot $\log \Gamma$ in units of  $(G/G_\mathrm{Q}) \phi_0$ versus the dimensionless $I_\mathrm{th}/I_\mathrm{f}$. Thermal activation with the effective temperature given by (\ref{eq:thermal}) would give a straight line (dashed lines in the plot). By virtue of our approach, the rates should exceed the quantum limit $\log \Gamma_{\hbar} \simeq (G/G_\mathrm{Q})\phi_0^2$.  This means that the rates should saturate at this value provided $I_\mathrm{f} \rightarrow 0$. For each choice of $S(\chi)$ we plot two curves corresponding to two possible signs of $V$ with respect to the current via the junction. For forward bias, the barrier is crossed when the fluctuating current is smaller than the average current. For backward bias, the barrier is crossed if the fluctuating current is bigger than the average value. The difference between two curves thus reflects the asymmetry of the current distribution with respect to the average current.

In Fig. 2, left panel, we present the results for a tunnel junction ($S_\mathrm{t}(\chi)= e^{i\chi}-1$) and a diffusive conductor ($S_\mathrm{d}(\chi)=(1/4)\mathrm{arccosh}^2(2e^{i\chi}-1)$)~\cite{dimanoisebook}. All curves approach the dashed thermal activation lines at  $I_\mathrm{f} \gg I_\mathrm{th}$. Since the tunnel junction is more noisy ($F=1$ versus $F=1/3$ for a diffusive conductor), it generally provides higher escape rates. However, the difference in functional form of the rates remains pronounced even upon rescaling with factor $3$. The most pronounced feature of the backward bias curves is a plateau at $I_\mathrm{f} \rightarrow I_\mathrm{th}$ with subsequent drop to very small escape rates $\simeq \Gamma_{\hbar}$ (beyond the vertical  scale of the plot). This is  because the current distribution is restricted: shot noise current is always of the same sign as the average current. 

\begin{figure}[b]
  \includegraphics[width=0.99\columnwidth]{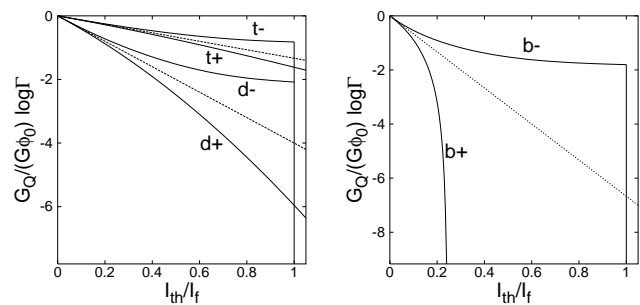}
  \caption{Escape rates versus $I_\mathrm{th}/I_\mathrm{f}$ for a tunnel (t), diffusive (d) and ballistic (b) mesoscopic conductor. ``+''/``-'' refers to forward/backward bias respectively. Dashed lines correspond to the rates due to Gaussian noise.}
\label{fig:rates}
\end{figure}

A quasi-ballistic conductor presents two peculiarities of this kind. We choose the transmissions of all channels to be the same, $T_0 =0.8$, $S_\mathrm{b}(\chi) = (1/T_0)\ln\left(1+T_0(e^{i\chi}-1)\right)$. In this case, the current distribution is restricted from both sides: the maximum current fluctuation can not exceed the ballistic limit $I_\mathrm{l} = I_\mathrm{f}/T_0$. From this we conclude that the barrier can not be crossed at forward bias if $I_\mathrm{th} > (1/T_0 -1)I_\mathrm{f}= 0.25 I_\mathrm{f}$, as seen in the right panel of Fig. 2. The rate becomes increasingly smaller upon approaching this threshold.

\begin{figure}
  \includegraphics[width=0.99\columnwidth]{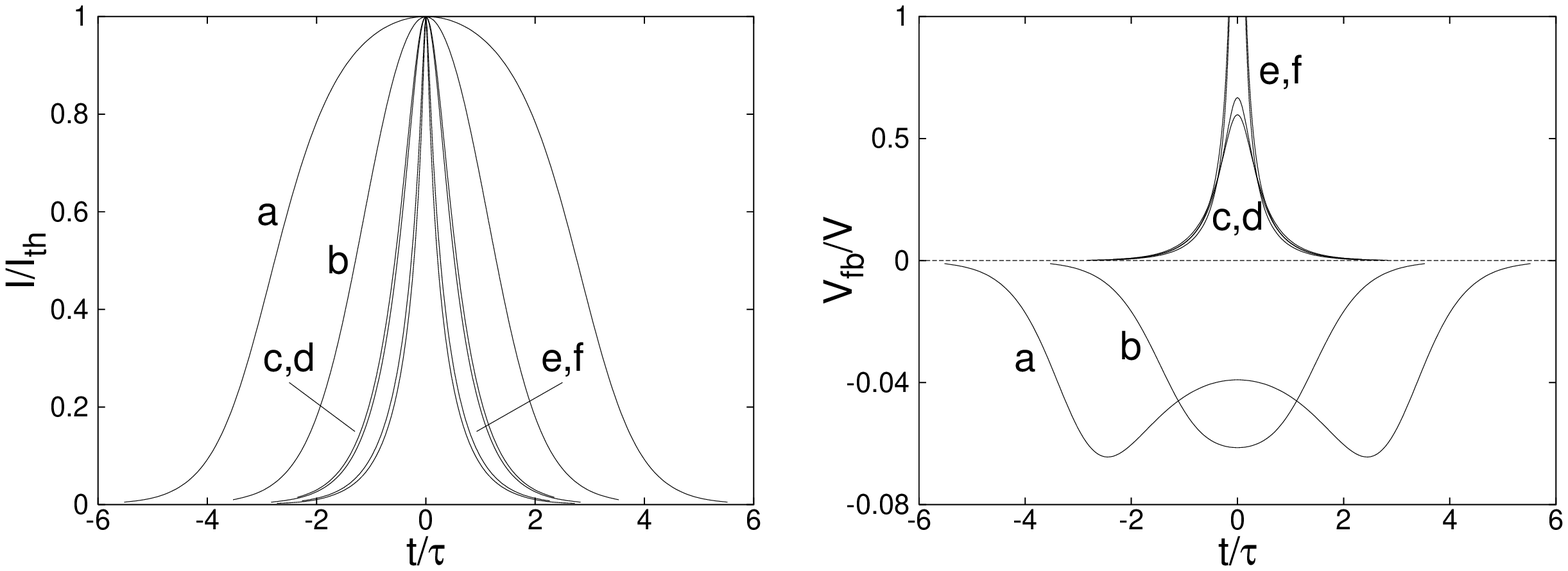}
  \caption{Optimal current fluctuations (left panel) and feedback voltage (right panel) versus time for different conductors and different values of $I_\mathrm{th}/I_\mathrm{f}$. Each line corresponds to one point on the curves in figure 2. Note the different voltage scales. The labels stand for (branch$\vert I_\mathrm{th}/I_\mathrm{f}$): (a) b+$\vert 1/5$, (b) b+$\vert 1/10$, (c) t-$\vert 1/3$, (d) d-$\vert 1/3$, (e) d-$\vert 2/3$, (f) t-$\vert 2/3$.}
\label{fig:fluctuations}
\end{figure}

There is an unambiguous correspondence between the rates as a function of $I_\mathrm{f}/I_\mathrm{th}$ and $S$, that is given by Eqs. (\ref{eq:actnsp}), (\ref{eq:phi}) and can be used to characterize the FCS from the rates measured. However, this relation is implicit and more complicated than that of an ideal detector. Apparently, this complication is due to the first and second factor mentioned in the introduction. To look at it in more detail, we compute the optimal current and voltage fluctuations that switch the detector.

The optimal current fluctuations are plotted in the left panel of Fig.~3 for different conductors and $I_\mathrm{f}$. The curves are symmetric owing to the symmetry of the cubic parabola potential. Common features are that they all reach the threshold current at maximum and their time spread is of the order of $\tau$. Still, the spread, shape, and most importantly, the integral of the current over time, varies significantly from curve to curve. This proves that the detector in use is dispersive and suffers from the first factor mentioned in the introduction.

The third factor is also in play. When $\phi$ moves, crossing the potential barrier, the resulting voltage changes the voltage drop over the normal conductor thereby affecting the current fluctuations in there. This feedback voltage $V_\mathrm{fb}$ is negative for forward bias and positive for negative one. We see from the evolution equations that
\begin{equation}
\frac{V_\mathrm{fb}}{V} \equiv -\frac{\hbar\dot{\phi}}{2e V}=
\frac{S(\chi(t))}{\chi(t)} \frac{\partial \chi}{\partial S(\chi(t))} -1,
\end{equation}  
so the change in the voltage drop across the junction is quite significant if $\chi \simeq 1$. We check that the negative feedback can never change the sign of the voltage for $S(\chi)$ in use. The right panel of Fig. 3 presents voltage fluctuations corresponding to the current fluctuations on the left panel. Interestingly, the positive feedback can be very big on the plateau at the backward bias (curves e, f). In this case, the detector seeks to optimize the rare fluctuation where almost no current is flowing in the normal conductor. The probability of such fluctuations is increased upon increasing the voltage drop over the conductor so that the detector provides the extra voltage required. Eventually, the feedback can be reduced with an extra resistive shunt over the Josephson junction. However, this would decrease $\tau$ and reduce the region of applicability of our results.

To conclude, we proved that Josephson junctions can be used as threshold detectors for non-Gaussian noise produced by coherent conductors. Our theoretical results facilitate a new type of electric noise measurement: direct measurement of the full distribution  of transferred charge.

We acknowledge fruitful discussions with P. Hadley, D. Esteve, M. H. Devoret and C. Markus. This work was supported by the Dutch Foundation for Fundamental Research on Matter (FOM). 

\bibliography{jjtdfcs}

\newpage

\end{document}